\documentclass[pra, twocolumn, showkeys, floatfix]{revtex4}
\usepackage{graphicx}
\usepackage{color}
\usepackage{amsmath, amsfonts, amssymb, bm}
\begin{document}
\title{Phenomenological rate formulas for over-barrier ionization of\\ 
hydrogen and helium atoms in strong constant electric fields}
\author{S.~Remme$^1$}
\author{A.~B.~Voitkiv$^1$}
\author{G.~Pretzler$^2$}
\author{C. M\"uller$^1$}
\affiliation{$^1$Institut f\"ur Theoretische Physik I, Heinrich-Heine-Universit\"at D\"usseldorf, 40225 D\"usseldorf, Germany \\
$^2$Institut f\"ur Laser- und Plasmaphysik, Heinrich-Heine-Universit\"at D\"usseldorf, 40225 Düsseldorf, Germany}
\date{\today}
\begin{abstract}
Nonrelativistic over-barrier ionization (OBI) of atoms in strong electric fields is studied, focussing on hydrogen and helium as concrete examples. Our goal is, on the one hand, to develop an intuitive physical picture behind established empirical formulas for the ionization rate. We show that the ionization rate in a near OBI regime can be modelled quantitatively by extending corresponding tunneling rates by the combined action of the Stark effect and a widened electron emission angle. On the other hand, we present analytical rate formulas in a far OBI regime which closely agree with available numerical data. In result, compact rate expressions describing OBI of hydrogen-like and helium atoms in a broad range of applied field strengths are obtained. They can be useful, for example, in numerical laser-plasma simulation codes to describe elementary ionization events.
\end{abstract}

\maketitle

\section{Introduction}
Ionization of atoms in strong electric fields is a fundamental
process of relevance in various areas of physics \cite{Review}. 
It constitutes, e.g., the starting point for high-harmonic and 
attosecond pulse generation \cite{HHG}
or advanced concepts of laser-plasma acceleration \cite{photocathode}.
Strong-field ionization is also of interest from a basic physical
perspective because, when the field is static or slowly varying,
the ionization possesses a manifestly nonperturbative character:
its rate depends exponentially on the inverse of the field strength.   
This renders strong-field ionization an important process to 
study nonperturbative physics in a quantum system.

As long as the applied field strength remains far below a certain threshold value, 
$F\ll F_c$, the ionization proceeds as a quantum tunneling process, 
where the electron escapes from the atom through the potential barrier 
that is formed by the superposition of the binding Coulomb potential and the applied 
external potential. Ionization in the tunneling regime is a well-understood
phenomenon and accurately described by the quasiclassical Ammosov–Delone–Krainov 
(ADK) \cite{ADK} (or Perelomov-Popov-Terent’ev \cite{PPT}) theory.

However, when the field amplitude exceeds the critical value $F_c$, 
the maximum of the deformed potential lies below the atomic binding 
energy and the electron may leave the atom above the barrier without 
the need to tunnel. The process then goes over from tunneling to 
over-barrier ionization (OBI). As a consequence, for $F\gtrsim F_c$, 
the ADK tunneling rate looses its applicability and substantially 
overestimates the ionization yield \cite{BauerMulser,
Scrinzi, TongLin, Lotstedt, Walker}.

While ionization rates in the tunneling regime have been derived in
analytical form, the OBI is more difficult to treat theoretically. 
Early attempts did not provide convincing results, as was shown in
\cite{BauerMulser} by comparison with 'exact' OBI rates that were 
obtained from solving the Schr\"odinger equation numerically in a 
wide range of applied field strengths. Apart from the possibility of 
fully numerical approaches \cite{BauerMulser, Scrinzi, TongLin, Lotstedt}, 
also compact empirical OBI rate formulas exist \cite{TongLin,Zhang} that 
represent extensions of the ADK tunneling rate. They have been obtained 
by parameter fits to numerical ionization data and achieve very good 
agreement for field strengths up to a few times the critical value.
A drawback of these successful empirical expressions is, however, that 
they lack a physical explanation or motivation for their functional form. 
It is one of the goals of the present paper to fill this gap by proposing 
a physically intuitive interpretation of these rate formulas in the near OBI regime.

The second goal of our paper is to present compact analytical formulas
for the ionization rate in the far OBI regime, where the applied field 
strength $F\gg F_c$ approaches or even exceeds the atomic field strength $F_a$.
On the one hand, a quadratic dependence of the rate on the applied field 
has been found around $F\sim 0.3F_a$ as an approximate fit to numerical data 
\cite{BauerMulser}, whereas on the other hand, a linear field dependence
has been predicted for $F\gtrsim F_a$ by a so-called motionless approximation
\cite{Kostyukov}. We will obtain a unified formula that is applicable for 
a broad range of field strengths and that is in good agreement with the 
available numerical data for hydrogen \cite{BauerMulser} and helium   
\cite{Scrinzi} in the far OBI regime. This formula can also be joined
with analytical expressions valid in the near OBI regime, so that the
full OBI range is covered.

Our consideration is restricted to the nonrelativistic domain. We point 
out that, very recently, a comprehensive theory for relativistic strong-field 
ionization including OBI has been developed \cite{Klaiber}. It relies on an 
adiabatic approximation for the transition amplitude and applies generalized eikonal
theory to describe the electron states. The ionization rate is then obtained 
by numerical evaluation of the corresponding amplitude.

The present paper is organized as follow. In Sec.~II we describe our 
approach to strong-field ionization in the near-OBI regime where $F\gtrsim F_c$.
It relies on physically motivated modifications to the ADK tunneling rate,
resulting from inclusion of the Stark effect and accounting for a widened
range of electron emission angles. The corresponding predictions will be
compared with established empirical OBI rate formulas \cite{TongLin,Zhang} 
and fully numerical data for hydrogen \cite{BauerMulser, Ivanov, Shabaev} and helium
\cite{Scrinzi}. Afterwards, the far-OBI regime of field strengths $F\gg F_c$ 
is considered in Sec.~III, where a novel fit formula is introduced that 
contains previously determined scaling laws \cite{BauerMulser,Kostyukov} 
approximately as limiting cases. Finally, our results in the near-OBI and 
far-OBI regimes are combined to provide compact empirical rate expressions 
for hydrogen and helium applicable in a broad range of field strengths. 
The conclusions from our considerations are given in Sec.~IV.

Atomic units (a.u.) are used throughout unless explicitly stated otherwise.
We note that one atomic unit of electric field strength corresponds to 5.14$\times$10$^9$~V/cm.

\section{Ionization in the near-OBI regime}
We consider ionization in a constant electric field $F$, pointing in negative $z$ direction. The corresponding interaction potential with the atomic electron reads $\hat{V} = -Fz$. The ADK formula of tunneling ionization has been derived for a general bound state with binding energy $-I_p$ and angular momentum quantum numbers $\ell$ and $m$ \cite{ADK, Madsen}. However, for our purposes it is sufficient to consider the special case of a $1s$ ground state, where the ADK formula reduces to \cite{TongLin,Madsen}
\begin{eqnarray}
\label{Landau}
W_{\rm TI}(F,\kappa) = \frac{C_a^2}{2\kappa^{2Z_c/\kappa-1}}\left(\frac{2\kappa^3}{F}\right)^{\!2Z_c/\kappa-1}\exp\left(-\frac{2\kappa^3}{3F}\right)\!,
\end{eqnarray}
with $\kappa = \sqrt{2I_p}$, the charge $Z_c$ of the atomic core, and an atom-specific parameter $C_a$ that reads $C_a=2\kappa^{3/2}$ for hydrogen-like systems. The rate $W_{\rm TI}$ features a characteristic dependence on $1/F$ in the exponential, which gives the rate a manifestly nonperturbative nature. Equation~\eqref{Landau} applies for field strengths $F\ll F_c$ far below the critical value
\begin{eqnarray}
\label{OBI-field}
F_c = \frac{\kappa^3}{16}\,,
\end{eqnarray}
where the Coulomb potential is suppressed all the way down to the (unperturbed) bound-state energy level \cite{F_c}. 

When the field is not too large ($F\lesssim 4F_c$), it is in principle possible to take the ADK tunneling formula as basis and modify it by suitable correction factors. By fitting to numerical ionization rates that have been obtained by the complex scaling method, Tong and Lin have proposed the following empirical OBI rate formula \cite{TongLin}:
\begin{eqnarray}
\label{Tong-Lin}
W_{\rm OBI}^{({\rm TL})}(F,\kappa) = W_{\rm TI}(F,\kappa)\,\exp\left(-\alpha\, \frac{Z_c^2}{I_p}\frac{F}{\kappa^3}\right),
\end{eqnarray}
with a single, atom-specific fit parameter $\alpha$. The latter has been determined for a number of atomic species in \cite{TongLin}. In particular, taking $Z_c=1$ and $\alpha = 6$ for hydrogen, this formula works well up to $F\approx 2.5F_c$. The correction factor depends on the properties of the considered atom and scales in the exponent linearly with the applied field.

By the same strategy, a three-parameter correction factor was obtained by Zhang et al. \cite{Zhang, sign}:
\begin{eqnarray}
\label{Zhang}
W_{\rm OBI}^{(\rm ZLL)}(F) = W_{\rm TI}(F,\kappa)\,\exp\left(a_1\frac{F^2}{F_c^2} + a_2\frac{F}{F_c} + a_3\right).
\end{eqnarray}
In addition to a linear term $\sim F$ in the exponent of the correction factor, it also contains a constant and a quadratic term $\sim F^2$. The enlarged number of fit parameters allows the authors to extend the applicability range of their OBI rate formula as compared with Eq.~\eqref{Tong-Lin}. For the case of hydrogen, Eq.~\eqref{Zhang} gives good results up to $F\approx 4.5F_c$, with the optimized values
$a_1=0.11714$, $a_2=-0.90933$ and $a_3=-0.06034$.

While both empirical formulas \eqref{Tong-Lin} and \eqref{Zhang} agree well with numerically determined  ionization rates, their drawback is to not provide a physical explanation for the functional form of the correction factors. In the following we present an intuitive, physically motivated procedure to obtain ionization rates in the near OBI regime by suitable modifications/extensions of the ADK formula \eqref{Landau}. Our approach relies on the inclusion of the Stark effect in both the bound-state energy and bound-state wave function along with a geometrically inspired rate averaging over the electron emission angles.

\subsection{Inclusion of Stark effect in binding energy}
The ADK theory relies on the assumption that the initial bound state remains unaltered in the presence of the external field. While this is justified for $F\ll F_c$, higher fields exert a sizeable influence on the bound electron by the Stark effect \cite{Walker, Klaiber, oriented1, oriented2, 1st-order}. In our treatment of near OBI we account for the Stark effect in the lowest (i.e. second) order of perturbation theory. The resulting shift of the binding energy in the $1s$ ground state of hydrogen reads
\begin{eqnarray}
\label{Stark-energy}
\varepsilon_S^{(2)} = \sum_{n>1}\frac{\big|\langle\psi_{n10}^{(0)}|\hat{V}|\psi_{100}^{(0)}\rangle\big|^2}{\varepsilon_1-\varepsilon_n} + \int\! d^3p\, \frac{\big|\langle\psi_{\vec p}^{(0)}|\hat{V}|\psi_{100}^{(0)}\rangle\big|^2}{\varepsilon_1-\varepsilon_p}.
\end{eqnarray}
Here, $\psi_{n\ell m}^{(0)}$ denotes the unperturbed bound state with quantum numbers $n,\ell,m$ and energy $\varepsilon_n$, whereas $\psi_{\vec p}^{(0)}$ is an unperturbed Coulomb wave with asymptotic momentum $\vec p$ and energy $\varepsilon_p = \frac{p^2}{2}$. The Stark shift energy in this case is known exactly; it is quadratic in the applied field and amounts to \cite{Landau} 
\begin{eqnarray}
\label{I_S}
\varepsilon_S^{(2)} = -I_S\ \ {\rm with}\ \ I_S = \frac{9}{4}\frac{F^2}{\kappa^4}\ .
\end{eqnarray}
One can already see here that the Stark shift leads to stronger binding of the ground state electron and, thus, to reduced ionization. As a side remark we note that by far the largest contribution to $I_S$ stems from the $2p_0$ state in Eq.~\eqref{Stark-energy} with almost 66\,\%; the rest comes in roughly equal shares from the higher lying bound states ($n\ge 3$) and the continuum states, respectively. The second-order perturbative prediction for the Stark shift \eqref{I_S} may be considered a reasonably good approximation up to field strengths $F\sim 0.1$\,a.u. ($\approx 1.5F_c$), where it underestimates the exact result by less than 25\%, according to the complex-scaling calculations in \cite{Ivanov,Shabaev}.

To extend the ADK rate formula into the near OBI regime, our first modification will be to replace the binding potential $I_p$ by $I_p+I_S$ in Eq.~\eqref{Landau}, both in the preexponential factor and the exponential. It is interesting to note that, by performing a Taylor expansion $(I_p+I_S)^{3/2} \approx I_p^{3/2} + \frac{3}{2}I_S I_p^{1/2}$ in the exponential of the resulting rate expression for $I_S\ll I_p$, one obtains
\begin{eqnarray}
\label{Stark1}
e^{-\frac{2\kappa^3}{3F}} \to 
e^{-\frac{2\kappa^3}{3F}}\exp\left( -\frac{9}{2}\frac{F}{\kappa^{3}} \right).
\end{eqnarray}
By comparison with Eq.~\eqref{Tong-Lin} we see that the exponential correction factor to the tunneling rate that results from inclusion of the (quadratic) Stark energy shift, has the same structure as the Tong-Lin factor, showing a linear dependence on $F$. The numerical prefactor amounts to $3/8$ of the corresponding prefactor in Eq.~\eqref{Tong-Lin}, which yields $W_{\rm OBI}^{\rm (TL)}/W_{\rm TI} = \exp(-12F/\kappa^3)$ for hydrogen. Thus, one may infer that a good portion of the correction factor of Tong and Lin can be associated with the Stark shift of the bound-state energy. For comparison we note that a factor of approx.~$-5$ instead of $-9/2$ in the exponent has been obtained by an adiabatic-eikonal treatment of strong-field ionization, where the Stark effect is included beyond perturbation theory (see Sec.\,III.B in \cite{Klaiber}).

It is worth noting at this point that the Stark effect has been included in the ADK tunneling rate to explain angular asymmetries arising in the strong-field ionization of oriented heteroatomic molecules \cite{oriented1, oriented2}. In this situation, also a linear Stark shift $\varepsilon_S^{(1)}$ occurs, since the considered hydrogen chloride (HCl) and carbonylsulphide (OCS) molecules possess permanent dipole moments. Depending on the relative orientation of the molecular axis to the applied field, the linear Stark shift either lowers or increases the ionization potential. This leads to correspondingly enhanced or reduced ionization yields, i.e.~to orientation-dependent asymmetries in the angular distributions of emitted photoelectrons. Also the quadratic Stark shift $\varepsilon_S^{(2)}$ was accounted for in the analysis of the experimental results \cite{oriented1, oriented2}.

While one might speculate that an observation like in Eq.~\eqref{Stark1} has motivated the form of the correction factor assumed by Tong and Lin, a link to the Stark effect is not mentioned in their paper \cite{TongLin}. Moreover, to the best of our knowledge, this connection has so far remained unnoticed in the literature. Being aware of it is important because sometimes the Stark shift up to second order is inserted via $I_p\to I_p-(\varepsilon_S^{(1)}+\varepsilon_S^{(2)})$ into the Tong-Lin formula \cite{molecules}. However, this can lead to a double counting of the quadratic Stark effect, because the Tong-Lin formula already contains this effect, according to the interpretation that Eq.~\eqref{Stark1} suggests.

\subsection{Inclusion of Stark effect in bound-state normalization}
The Stark effect also affects the wave function $\psi_{100}^{(0)}$ of the bound state, leading to a polarization of the atom along the field direction. Two contributions to the atomic polarization can be distinguished \cite{Klaiber}: (i) the shift of the bound electron state towards the 'tunnel exit'; this effect, which leads to {\it enhanced} ionization, is implicitly accounted for in the ADK formula by a corresponding shift of the matching point between the unperturbed bound state and the continuum state of the electron \cite{ADK, Landau, Madsen}. (ii) the bound state distortion, that causes a {\it reduction} of the ionization rate; since the ADK rate does not include this effect, it needs to be modified by a corresponding correction factor in the OBI regime. Below we shall argue how an approximate version of such a correction factor may be obtained within perturbation theory.

The first-order correction of the unperturbed ground state is given by 
\begin{eqnarray*}
\psi_{100}^{(1)}\! =\! \sum_{n>1}\frac{\langle\psi_{n10}^{(0)}|\hat{V}|\psi_{100}^{(0)}\rangle}{\varepsilon_1-\varepsilon_n}\psi_{n10}^{(0)} +\! \int\! d^3p\,\frac{\langle\psi_{\vec{p}}^{(0)}|\hat{V}|\psi_{100}^{(0)}\rangle}{\varepsilon_1-\varepsilon_p} \psi_{\vec{p}}^{(0)}
\end{eqnarray*}
and, thus, orthogonal to the unperturbed ground state. In contrast, the second-order correction $\psi_{100}^{(2)}$ contains a contribution proportional to $\psi_{100}^{(0)}$, leading to a non-zero overlap $\langle \psi_{100}^{(2)}|\psi_{100}^{(0)}\rangle = -\frac{1}{2}\beta F^2$ with \cite{Landau}
\begin{eqnarray}
\beta = \sum_{n>1}\left|\frac{\langle\psi_{n10}^{(0)}|\hat{z}|\psi_{100}^{(0)}\rangle}{\varepsilon_1-\varepsilon_n}\right|^2 +\! \int\! d^3p\,\left|\frac{\langle\psi_{\vec{p}}^{(0)}|\hat{z}|\psi_{100}^{(0)}\rangle}{\varepsilon_1-\varepsilon_p}\right|^2.
\end{eqnarray}
The numerical value of this parameter turns out as $\beta\approx 4.912$. The correspondingly perturbed state $\psi_{100} = \psi_{100}^{(0)}+\psi_{100}^{(1)}+\psi_{100}^{(2)}$ (that is normalized up to order $\mathcal{O}(F^2)$) hence contains the portion $\psi_{100} = N \psi_{100}^{(0)}+\ldots$ proportional to the unperturbed ground state, where 
\begin{equation}
N = 1-\frac{\beta}{2} F^2
\label{norm}
\end{equation} 
represents an adjusted 'normalization constant' for the $\psi_{100}^{(0)}$ contribution.
Our second modification of the ADK formula due to the Stark effect shall be to multiply Eq.~\eqref{Landau} by $|N|^2\approx 1-\beta F^2$, since the ground state $\psi_{100}^{(0)}$ enters quadratically in the ADK ionization rate. 

We emphasize that all the other contributions to the perturbed ground state $\psi_{100}$ are dropped in our consideration. Essentially, this simplifying step can only be justified {\it a posteriori} by a remarkably good agreement of the predictions from our phenomenological treatment with accurate numerical ionization rates (see Sec.\,III). Additionally, it might be argued that the contributions from the excited bound and continuum states to $\psi_{100}$ are mainly responsible for the polarization shift of the ground state, which is already contained in the ADK rate.

The modified normalization constant \eqref{norm} may be interpreted as reflecting (part of) the distortion of the initial-state wave function by the Stark effect. For comparison we note that in Sec.\,III.B of \cite{Klaiber}, a corrected normalization constant of $1-\frac{31}{4}F^2$ has been found, which possesses the same structure as Eq.~\eqref{norm}. In line with the results of this recent study, both the Stark shift of the binding energy \eqref{I_S} and the Stark distortion of the wave function \eqref{norm} lead to reductions of the ionization rate. 

The alteration of the state normalization can be linked, at least qualitatively, to the appearance of a quadratic term in the correction factor of Zhang et al. in Eq.~\eqref{Zhang} because, provided that $\beta F^2\ll 1$ holds, one may write approximately $|N|^2 \approx \exp\big(-\beta F^2\big)$.

We conclude Secs.\,II.A and B by noting that an asymptotic perturbative expansion of the ionization rate for hydrogen in powers of the field has obtained a correction factor up to quadratic order of $1-\frac{53}{12}F-\frac{3221}{8\cdot 36}F^2$ \cite{Damburg}. We note that the numerical factor in the linear term is very close to the $9/2$ in Eq.~\eqref{Stark1}. An extension of this weak-field asymptotic theory has been provided via Siegert states \cite{1st-order}; the corresponding results are applicable to a larger class of atoms and field strengths approaching the boundary between tunneling and over-barrier ionization.

\subsection{Accounting for widened range of emission angles}
\begin{figure}[b]  
\begin{center}
\includegraphics[width=0.35\textwidth]{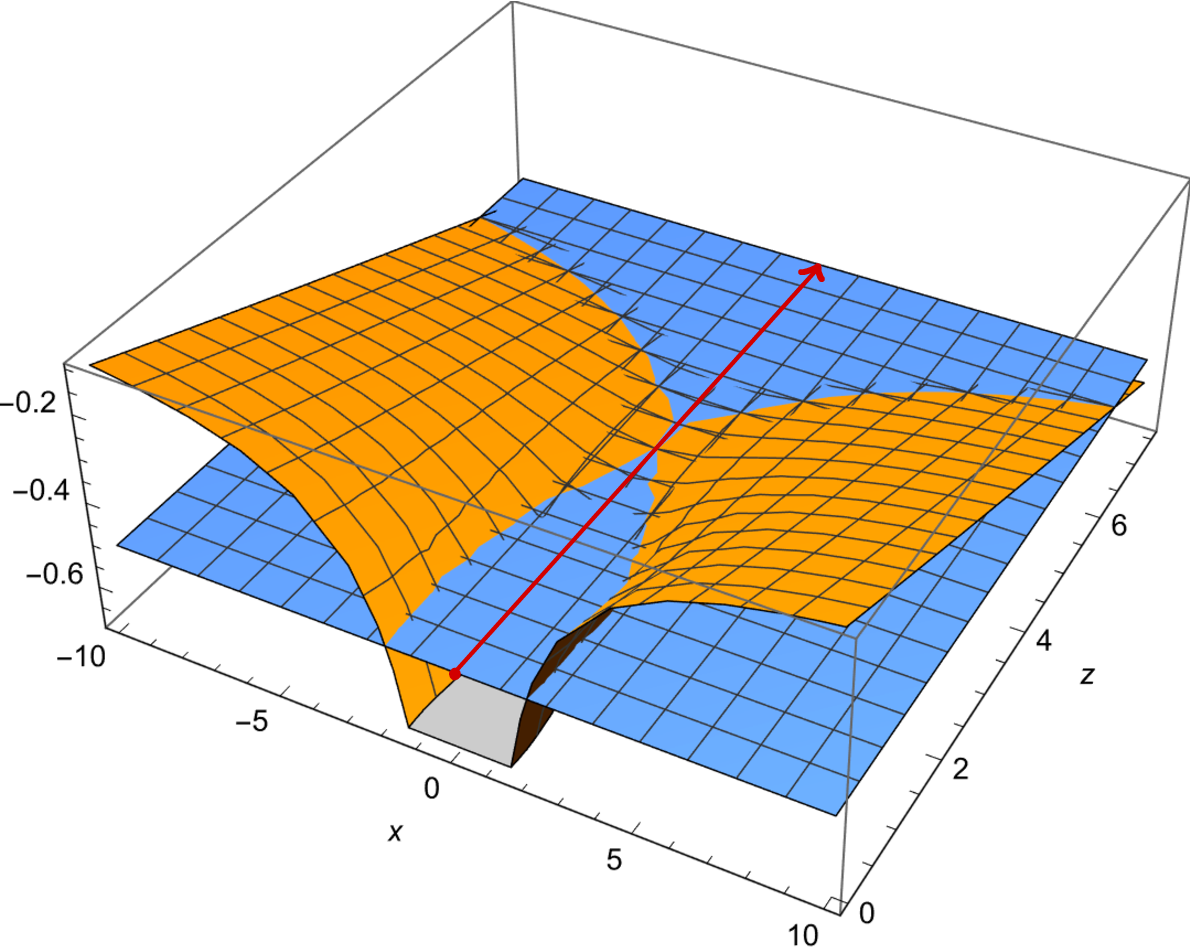}
\includegraphics[width=0.35\textwidth]{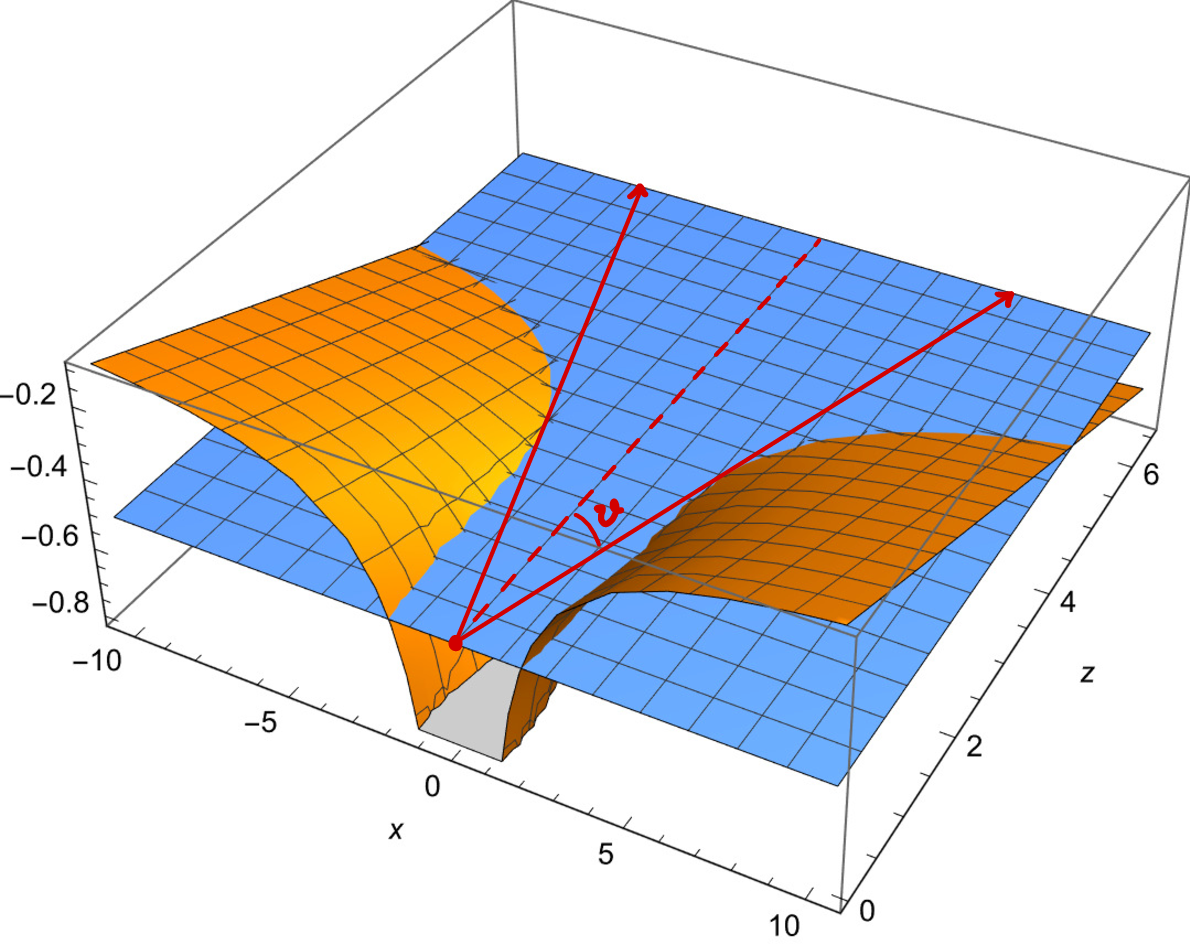}
\includegraphics[width=0.35\textwidth]{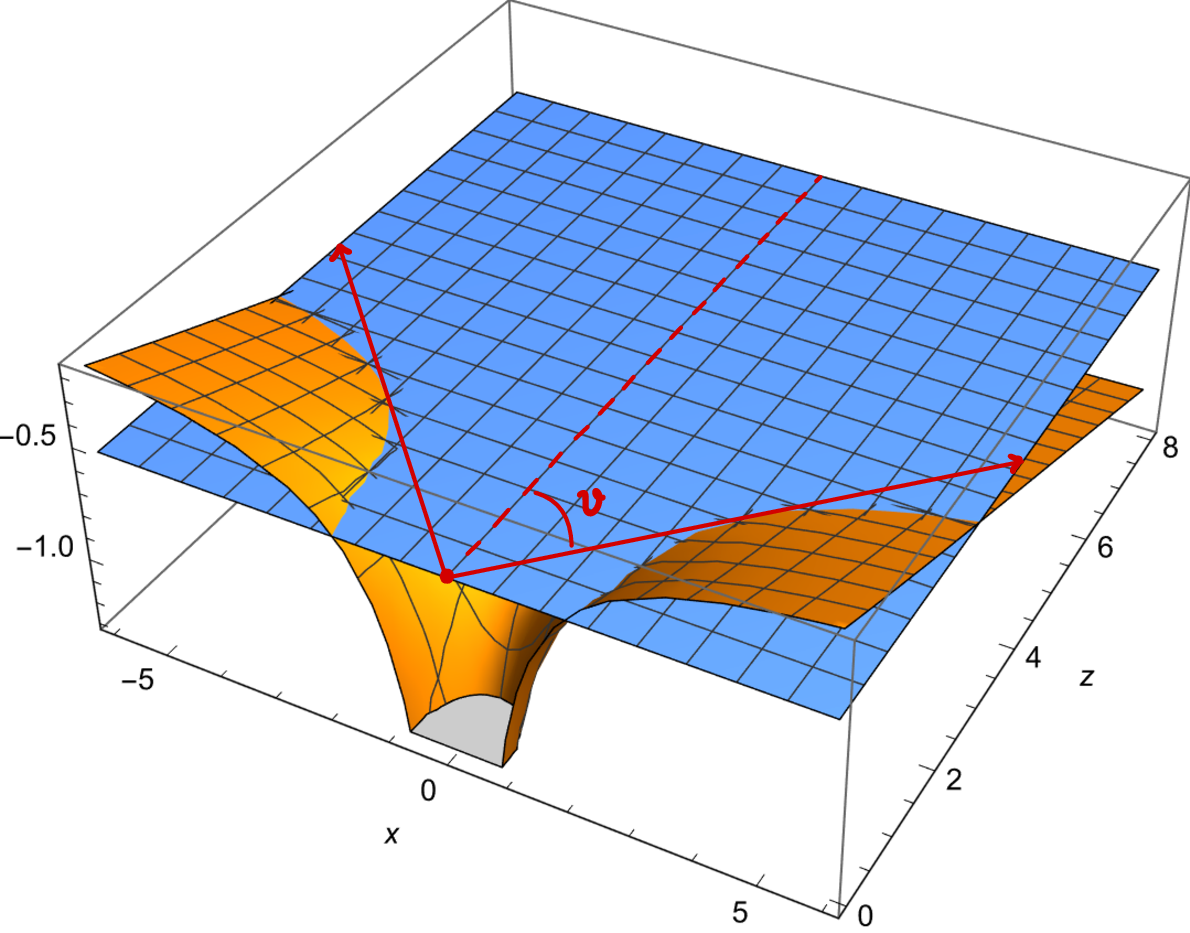}
\end{center}
\caption{Total potential $\hat{V}_{\rm tot} = -Fr\cos\vartheta - \frac{1}{r}$, consisting of the external field potential and the atomic Coulomb potential of hydrogen (curved orange plane). The horizontal plane in blue indicates the location of the Stark-shifted bound-state energy $-(I_p+I_S)$. The external field is taken as $F = 1.04F_c$ in the top panel, $F=1.5F_c$ in the middle panel, and $F=2F_c$ in the bottom panel.}
\label{fig-potential}
\end{figure}

With the two Stark-effect modifications from Sec.\,II.A and II.B included, our preliminary expression for the ionization rate in the near OBI regime reads
\begin{eqnarray}
\label{rate-with-Stark}
\tilde{W}_{\rm nOBI}(F,\kappa') = |N|^2\, W_{\rm TI}(F,\kappa')
\end{eqnarray}
with $\kappa' = \sqrt{2(I_p+I_S)}$.
The final modification that we are going to carry out to the ADK formula \eqref{Landau} is geometrically motivated. In a tunneling description, the electron leaves the atom -- essentially -- along the $z$ direction, i.e. opposite to the applied field \cite{Landau, Madsen}. This is reasonable because in this direction the tunnel is shortest and the potential barrier lowest. The other emission directions are subject to even stronger exponential suppressions and may therefore be neglected. Conversely, in the OBI regime, the 'door is open' for the electron since the potential is suppressed completely. This holds not only for the direction along the field axis, but for a range of emission angles, as is illustrated in Fig.~\ref{fig-potential}. 

We can determine the maximum angle $\vartheta_{\rm max}$ under which the electron can leave the atom above the barrier from the total potential $\hat{V}_{\rm tot}(r,\vartheta) = -Fr\cos\vartheta - \frac{\kappa}{r}$ by demanding that its maximum value with respect to the radial coordinate equals the Stark-shifted ground-state energy. This leads to the relation $2\sqrt{\kappa F\cos\vartheta_{\rm max}} = I_p+I_S$, from which
\begin{eqnarray}
\label{theta-max}
\vartheta_{\rm max} = \arccos\left[\frac{F_c}{F}\left(1+\frac{I_S}{I_p}\right)^{\!2}\right]
\end{eqnarray} 
follows. Thus, when emitted within a cone of opening angle $\vartheta_{\rm max}$ around the $z$ axis (about which the problem is symmetric), the electron is ionized above the barrier.
Since the emission directions within this cone are not suppressed by the need to tunnel, all of them may contribute to ionization. The electron thus has a 'free choice' of pathways and since one does not know which way it goes, it appears reasonable to take an average over the relevant range of emission angles, according to
\begin{eqnarray}
\label{averaged-rate}
W_{\rm nOBI}(F,\kappa) = \frac{1}{\vartheta_{\rm max}}
\int_0^{\vartheta_{\rm max}}\tilde{W}_{\rm nOBI}(F\cos\vartheta,\kappa')\, d\vartheta\,.
\end{eqnarray}
This angular averaging can be applied for field strengths $F\ge 0.065\,{\rm a.u.} = 1.04 F_c$, where the argument of the arccosine in Eq.~\eqref{theta-max} is $\le 1$. The corresponding integral in Eq.~\eqref{averaged-rate} is evaluated numerically. 
We note that the maximal opening angle $\vartheta_{\rm max}\to 0$ when the applied field strength approaches to $\approx 1.04F_c$ from above. In this limit, the integrand in Eq.~\eqref{averaged-rate} is approximately constant and the equation goes over to $W_{\rm nOBI}(F,\kappa) = \tilde{W}_{\rm nOBI}(F,\kappa')$, so that a smooth transition towards smaller fields $F<1.04F_c$ is guaranteed.

In order to properly interpret the angular average in Eq.~\eqref{averaged-rate} it is worth recalling that, in the derivation of the ADK formula \eqref{Landau}, the $z$-component of the electron current density -- which lies antiparallel to the field direction -- is transformed into an ionization rate by an area integral over a transverse plane \cite{Landau, Madsen}. Accordingly, since the modified tunneling rates $\tilde{W}_{\rm nOBI}(F\cos\vartheta,\kappa')$ on the right-hand side of Eq.~\eqref{averaged-rate} rely on Eq.~\eqref{Landau}, they contain this area integration. However, instead of an electron flux in $z$-direction (corresponding to $\vartheta=0$), they refer to electron emission under the angle $\vartheta$ which may vary from zero up to $\vartheta_{\rm max}$. In this way, a broader range of electron emission directions is taken into account.

\subsection{Near-OBI results for hydrogen and helium}
In the following we will compare the predictions from our rate formulas \eqref{rate-with-Stark} and \eqref{averaged-rate}, including the Stark effect and the widened emission range, with other empirical rates \cite{TongLin, Zhang} and with fully numerical results \cite{Scrinzi, Shabaev}. 

\begin{figure}[t]  
\vspace{-0.25cm}
\begin{center}
\includegraphics[width=0.45\textwidth]{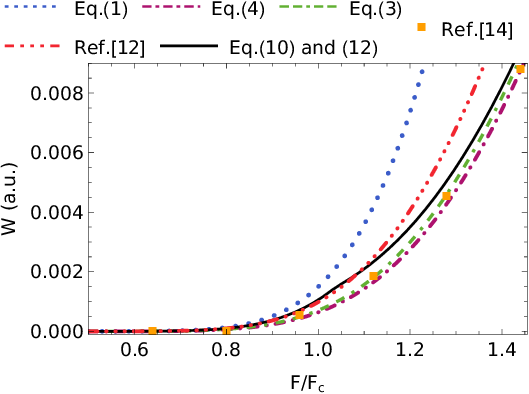}
\end{center}
\vspace{-0.5cm} 
\caption{Ionization rates for atomic hydrogen from the ground state in a near-OBI regime of field strengths around $F_c$. The dotted blue line shows the ADK tunneling rate \eqref{Landau}, the dash-dotted purple line the empirical formula \eqref{Zhang} by Zhang et al. \cite{Zhang}, the double-dash-dotted green line the empirical formula \eqref{Tong-Lin} by Tong and Lin \cite{TongLin}, and the dash-triple-dotted red line is based on Eq.~(A2) of Klaiber et al. \cite{Klaiber}. The orange squares are numerical results by Maltsev et al.~\cite{Shabaev}. The black solid line shows our Eq.~\eqref{rate-with-Stark} for $F\le 1.04 F_c$ and our Eq.~\eqref{averaged-rate} for $F\ge 1.04F_c$, with the latter involving the angular averaging.}
\label{nOBI-H-1}
\end{figure}

Figure~\ref{nOBI-H-1} shows the ionization rate of hydrogen in a near-OBI regime around the critical field strength $F_c$. The ADK tunneling rate \eqref{Landau}, shown by the dotted blue curve, is seen to  significantly overestimate the ionization. Our rate prediction (solid black line) lies slightly above the empirical rates \eqref{Tong-Lin} and \eqref{Zhang}, which agree well with the 'exact' rate resulting from fully numerical complex-scaling calculations \cite{Shabaev}. The difference between our result and these other predictions is largest close to $F\approx F_c$; it starts to improve significantly when the angular averaging is included for $F\ge 1.04F_c$. For the highest field strengths shown in the figure, our rate reaches close agreement with the predictions from Eqs.~\eqref{Tong-Lin}, \eqref{Zhang} and Ref.\,\cite{Shabaev}. The figure also shows by the dash-triple-dotted red line an analytical approximation from Ref.\,\cite{Klaiber}, based on Eq.~(A2) therein for the ionization amplitude (including a factor $\frac{1}{2}$); at higher field values, the corresponding rate lies slightly above the other OBI predictions. 

\begin{table}[b]\label{table1}
\begin{tabular}{c|c|c|c}
\hline
\hline
 & $\tilde{a}_1$ & $\tilde{a}_2$ & $\tilde{a}_3$ \\
\hline
hydrogen\ \ &\ \ 0.047769\ \ &\ \ -0.745282\ \ &\ \ -0.114022\ \ \\
\hline
helium\ \ &\ \ 0.042177\ \ &\ \ -0.762430\ \ &\ \ 0.039149\ \ \\
\hline
\hline
\end{tabular}
\caption{Optimized parameters to fit the near-OBI rate from Eq.~\eqref{averaged-rate} to the analytical form \eqref{nOBI-fit}.}
\end{table}

\begin{figure}[t]
\vspace{-0.25cm}
\begin{center}
\includegraphics[width=0.45\textwidth]{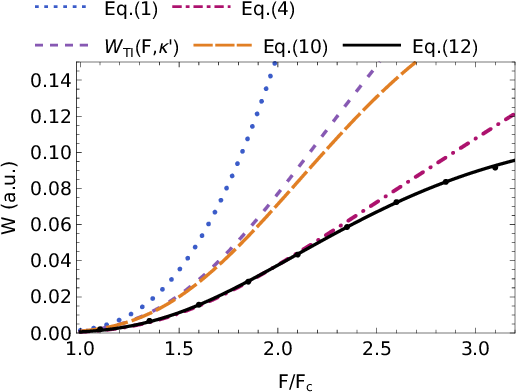}
\end{center}
\vspace{-0.5cm} 
\caption{Ionization rates for atomic hydrogen from the ground state in a near-OBI regime of field strengths above $F_c$. The dotted blue line shows the prediction of the ADK tunneling rate \eqref{Landau}, the short-dashed violet line includes the rate modification from the Stark-shifted binding energy, the long-dashed orange line includes in addition the Stark-modification of the wave-function normalization, and the black marker points show the results from Eq.~\eqref{averaged-rate}. The solid black line connecting the marker points shows the fit formula \eqref{nOBI-fit}. The dash-dotted purple line results from the empirical formula~\eqref{Zhang}.}
\label{nOBI-H-2}
\end{figure}

Figure~\ref{nOBI-H-2} focusses on the near-OBI regime above $F_c$. It shows the empirical rate \eqref{Zhang} as a (dash-dotted purple) reference line, since the numerical results of \cite{Shabaev} are restricted to $F\le 0.1\,{\rm a.u.}\approx 1.5F_c$. When the ADK tunneling rate $W_{\rm TI}(F,\kappa)$  (dotted blue line) is amended to $W_{\rm TI}(F,\kappa')$ by inclusion of the Stark energy shift, the short-dashed violet curve results. It lies already much closer to the reference rate. Additional account of the Stark effect in the wave function normalization leads to a further approach, as the long-dashed orange line displays. Inclusion of the widened angular emission range according to Eq.~\eqref{averaged-rate} finally leads to very good agreement between our rate prediction (black marker points) with the empirical rate \eqref{Zhang} by Zhang et al.~up to field strengths $F\approx 2.5F_c$. The black solid line in Fig.~\ref{nOBI-H-2} was obtained as an analytical fit of the form [cf.~also Eq.~\eqref{Zhang}]
\begin{eqnarray}
\label{nOBI-fit}
W_{\rm nOBI}^{(\rm fit)} = W_{\rm TI}(F,\kappa)\,
\exp\left(\tilde{a}_1\frac{F^2}{F_c^2} + \tilde{a}_2\frac{F}{F_c} + \tilde{a}_3\right)
\end{eqnarray}
to our results from Eq.~\eqref{averaged-rate}. The values of the fit parameters that we obtained, are given in Table I. We may thus conclude that our rate formula \eqref{averaged-rate}, that was obtained by physically motivated modifications to the ADK tunneling rate, has a similar range of applicability for OBI of hydrogen as the empirical rate \eqref{Tong-Lin} by Tong and Lin.
We note that, interestingly, the upper boundary of the applicability range of about $2.5F_c\approx 0.16$\,a.u. lies close to the alternative value of 0.146\,a.u.~of the critical OBI field for hydrogen obtained in \cite{F_c}. This coincidence might explain why the ionization rate in this regime can be described by modified tunneling-like formulas. We furthermore note that, in the limit of low fields $F\ll F_c$, the modified rate formula in Eq.~\eqref{nOBI-fit} approximately goes over to the ADK tunneling rate $W_{\rm TI}(F,\kappa)$, because the absolute value of the fit parameter $\tilde{a}_3$ is small, $|\tilde{a}_3|\ll 1$.

Our results for hydrogen carry over to OBI of hydrogen-like ions with nuclear charge $Z$ by virtue of a scaling transformation. The rate for ionization of an ion with charge $Z$ in a field of strength $F$ is obtained from the rate for ionization of hydrogen by evaluating the latter at the scaled field $F/Z^3$ and multiplying the result with $Z^2$. Since our treatment is nonrelativistic, the nuclear charge is restricted to values of about $Z\lesssim 20$ where relativistic effects in the ion may be neglected.

\begin{figure}[t]
\vspace{-0.25cm}
\begin{center}
\includegraphics[width=0.45\textwidth]{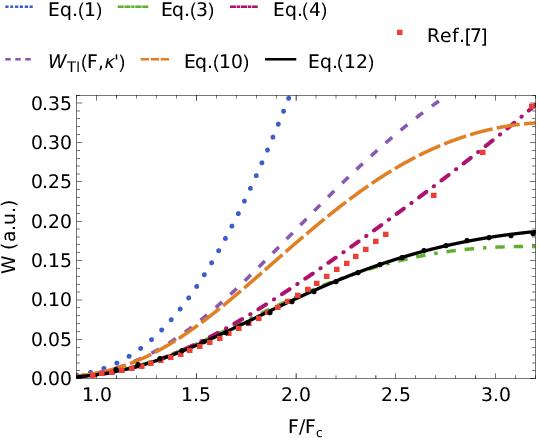}
\end{center}
\vspace{-0.5cm} 
\caption{Ionization rates for helium from the ground state in a near-OBI regime of field strengths above $F_c=0.204$\,a.u. The dotted blue line shows the ADK tunneling rate, the short-dashed violet line includes the rate modification from the Stark-shifted binding energy, the long-dashed orange line includes in addition the Stark-modification of the wave-function normalization, and the black marker points show the results from Eq.~\eqref{averaged-rate}. The solid black line connecting the marker points shows the fit formula \eqref{nOBI-fit}. The double-dash-dotted green line is the empirical formula \eqref{Tong-Lin} by Tong and Lin \cite{TongLin} (with $\alpha = 7$), the dash-dotted purple line is the empirical formula \eqref{Zhang} by Zhang et al. \cite{Zhang}, and the red marker squares are the numerical results by Scrinzi et al.~\cite{Scrinzi}.}
\label{nOBI-He}
\end{figure}

By some suitable adjustments, we can apply our empirical approach also to OBI of helium. A comparison of corresponding near-OBI rates obtained by different methods is shown in Fig.~\ref{nOBI-He}. Since the helium ground state contains two $1s$ electrons, the ADK rate has been multiplied by a factor of 2 and the atom-specific parameter in Eq.~\eqref{Landau} set to $C_a=2.67$ \cite{ADK,Tong-2002}. The critical field in case of helium is taken as $F_c = 0.204$\,a.u. \cite{TongLin, Lotstedt}.

In order to catch the main physical properties of helium within our effective one-electron model, we shall use effective nuclear charges. Into the ADK tunneling rate \eqref{Landau}, the binding potential $I_p = 0.904$\,a.u. enters, corresponding to $\kappa\approx 1.345$. To estimate the Stark effect in helium, we use an effective nuclear charge $Z_S\approx 1.44$ that results from the $1s$-$2p$ transition energy. Our choice is motivated by the fact, that this transition gives by far the largest contribution to the Stark shift of the binding energy [see the remark below Eq.~\eqref{I_S}]. This way, we obtain $I_S \approx 0.523 F^2$ \cite{polarizability} from Eq.~\eqref{I_S} and $\beta\approx 0.551$ in the normalization constant \eqref{norm}. 

As Fig.~\ref{nOBI-He} illustrates, by successively including the Stark shift to the binding energy (short-dashed violet curve), the Stark effect in the wave function normalization (long-dashed orange curve), and the widened emission range (gray marker points, connected by solid gray line), the rate prediction approaches more and more to the 'exact' ionization rate. The latter has been obtained through fully numerical complex-scaling calculations by Scrinzi et al.~\cite{Scrinzi}. We note that these numerical rates have recently been confirmed by multiconfiguration time-dependent Hartree-Fock computations \cite{Lotstedt}.

\section{Ionization of hydrogen and helium in the far-OBI regime}
When the external field strength $F$ increases beyond a few times the critical field $F_c$, the empirical Eqs.~\eqref{Tong-Lin}, \eqref{Zhang} and \eqref{averaged-rate} are not applicable anymore. This is expected because the ionization mechanism then differs too strongly from tunneling, so that rate expressions obtained by modifications of the ADK formula do not provide a suitable description in a far-OBI regime. Instead, the rate dependence has been found to change qualitatively. For OBI of hydrogen in the range $0.15\le F/F_a\le 0.5$, an approximately quadratic field dependence 
\begin{eqnarray}
\label{Bauer}
W_{\rm OBI}^{\rm (BM)}\approx 2.4 F^2
\end{eqnarray}
was obtained by Bauer and Mulser through fitting to ionization rates from a fully numerical solution of the Schr\"odinger equation \cite{BauerMulser}. For even higher fields up to about $4F_c$ their numerical results indicated a flattening of the quadratic dependence. Based on a so-called motionless approximation, Kostyukov and Golovanov have predicted a linear increase of the rate \cite{Kostyukov}, according to
\begin{eqnarray}
\label{Kostyukov}
W_{\rm OBI}^{\rm (KG)}\approx 0.8F\,.
\end{eqnarray}
This formula is supposed to be applicable for $F\gg 3F_c$, reaching up to few times the atomic field $F_a$.

Guided by these findings, we propose the following functional form to model the ionization rate in the far-OBI regime:
\begin{eqnarray}
\label{fOBI-rate}
W_{\rm fOBI} = \frac{aF^b}{c+F^d}\,.
\end{eqnarray}
It contains four fit parameters that will be adjusted to existing numerical OBI rates. The form of Eq.~\eqref{fOBI-rate} implies that, for $b\approx 2$ and $d\approx 1$, the field dependence of the rate resembles the quadratic scaling of Eq.~\eqref{Bauer} for moderate OBI field strengths, while for very large fields it approaches approximately to a linear growth as in Eq.~\eqref{Kostyukov}. 

In order to determine the fit parameters for hydrogen, we use the predictions from the empirical formula \eqref{Zhang} by Zhang et al.~for $F=3.5F_c$ and $4F_c$, and the numerical rates computed by Bauer and Mulser for higher field strengths $0.6\lesssim F/F_a\lesssim 4$ (see Fig.~6 in \cite{BauerMulser}). As a result, we obtain the values provided in Table II. Accordingly, for moderate field strengths of $F\approx 0.3F_a$ and below, the rate \eqref{fOBI-rate} approximately scales as $W_{\rm fOBI}\sim F^{b_{\rm H}} \approx F^{2.27}$, resembling the quadratic dependence of Eq.~\eqref{Bauer}. For large fields $F\gtrsim F_a$ one obtains instead a nearly linear rate growth $W_{\rm fOBI}\sim F^{b_{\rm H}-d_{\rm H}} \approx F^{0.83}$ like in Eq.~\eqref{Kostyukov}. 

\begin{table}[b]\label{table2}
\begin{tabular}{l|c|c|c|c}
\hline
\hline
 & $a$ & $b$ & $c$ & $d$ \\
\hline
hydrogen\ \ &\ \ 2.12576\ \ &\ \ 2.27298\ \ &\ \ 0.39435\ \ &\ \ 1.44093\ \ \\
\hline
helium\ \ &\ \ 1.11521\ \ &\ \ 3.46003\ \ &\ \ 0.35354\ \ &\ \ 2.23355\ \ \\
\hline
\hline
\end{tabular}
\caption{Optimized parameters to fit the far-OBI rate to the analytical form given in Eq.~\eqref{fOBI-rate}.}
\end{table}

\begin{figure}[t]  
\vspace{-0.25cm}
\begin{center}
\includegraphics[width=0.45\textwidth]{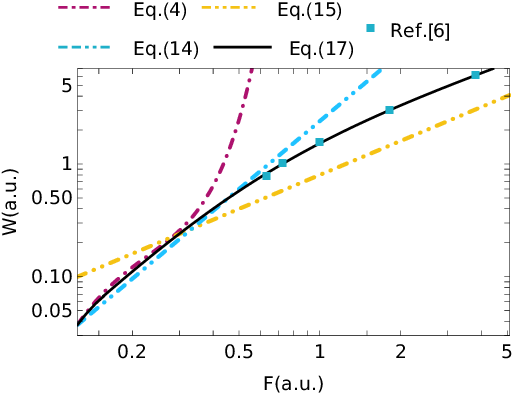}
\end{center}
\vspace{-0.5cm} 
\caption{OBI rates for atomic hydrogen in a wide range of field strengths, 
including the far-OBI regime. The solid black line shows our combined rate 
formula \eqref{full-rate}. For comparison, the double-dash-double-dotted cyan 
line is the quadratic rate \eqref{Bauer} by Bauer and Mulser, that was obtained 
in the range $0.15\,{\rm a.u.}\le F\le 0.5\,{\rm a.u.}$; the cyan squares at higher 
field strengths are read off from Fig.~6 in \cite{BauerMulser}. The dash-double-dotted 
yellow line is the linear rate \eqref{Kostyukov} predicted by Kostyukov and Golovanov
\cite{Kostyukov}, whereas the dash-dotted purple line is the rate \eqref{Zhang} by 
Zhang et al.~\cite{Zhang}.}
\label{OBI-H}
\end{figure}

Our prediction for the OBI rate of hydrogen is shown by the solid black curve in Fig.~\ref{OBI-H}.
The curve represents the piecewise defined function
\begin{eqnarray}
\label{full-rate}
W_{\rm OBI} = \left\{ \begin{array}{ll}
W_{\rm nOBI}^{\rm (fit)} & {\rm for}\ F\le F_{\rm tr} \\
W_{\rm fOBI} & {\rm for}\ F > F_{\rm tr} 
\end{array} \right.
\end{eqnarray}
which combines Eq.~\eqref{nOBI-fit} with Eq.~\eqref{fOBI-rate} and, thus, extends from the near-OBI to the far-OBI regime. The 'transition field strength' in case of hydrogen is chosen as $F_{\rm tr}=2.5F_c$, since Eq.~\eqref{nOBI-fit} works well up to this point (see Fig.~\ref{nOBI-H-2}). For $F\ge 2.5 F_c\approx 0.16$\,a.u., the curve is obtained from Eq.~\eqref{fOBI-rate} with the fit parameters given in Table II. It runs very closely to the empirical rate of Zhang et al.~(dash-dotted purple curve) up to $F\approx 0.3$\,a.u. and to the rate formula \eqref{Bauer} of Bauer and Mulser up to $F\approx 0.5$\,a.u. Our rate prediction also matches well the numerical data of Bauer and Mulser (cyan marker points) at higher fields, where their quadratic rate formula \eqref{Bauer} starts to deviate. In this region, our rate runs almost parallely to---but significantly higher than---the result of Eq.~\eqref{Kostyukov} by Kostyukov and Golovanov. Overall, the combined formula \eqref{full-rate} accomplishes very satisfactory OBI rate predictions in a broad range of field strengths.

\medskip

\begin{figure}[t]
\begin{center}
\includegraphics[width=0.45\textwidth]{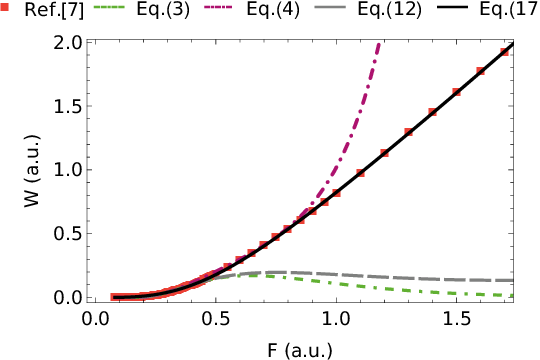}
\end{center}
\vspace{-0.5cm} 
\caption{OBI rates for helium, spanning the range from near- to far-OBI.
The black solid curve shows our combined rate formula \eqref{full-rate},
while the red marker squares are the numerical results from Scrinzi et 
al.~\cite{Scrinzi}. For comparison, the dash-dotted purple line is the 
rate \eqref{Zhang} by Zhang et al., the dash-dotted green curve is the 
rate \eqref{Tong-Lin} by Tong and Lin, whereas the long-dashed gray rate 
refers to our near-OBI formula \eqref{averaged-rate}, 
which is applicable up to $F\approx 2.05F_c\approx 0.42$\,a.u.}
\label{OBI-He}
\end{figure}

Our results from Eq.~\eqref{full-rate} for OBI of helium are shown by the black solid curve in Fig.~\ref{OBI-He}. Here, we take the transition point between near-OBI and far-OBI at $F_{\rm tr} = 2.05F_c \approx 0.42$\,a.u., in accordance with Fig.~\ref{nOBI-He}. The fit parameters entering $W_{\rm fOBI}$ are listed in Tab.~II; they have been obtained based on the numerical results of Scrinzi et al.~\cite{Scrinzi}. Our corresponding analytical rate prediction of the form \eqref{fOBI-rate} reaches very good agreement with these 'exact' numerical data in the full range of considered field strengths. For very large fields $F > 1$\,a.u., the rate $W_{\rm fOBI}\sim F^{b_{\rm He}-d_{\rm He}} \approx F^{1.23}$ scales approximately linearly, in accordance with Eq.~\eqref{Kostyukov}. However, for helium the rate grows slightly faster than linearly, whereas for hydrogen a slightly sub-linear growth $\sim F^{0.83}$ was obtained above. In an intermediate regime of field strengths around $F\sim 0.5$\,a.u. (i.e. $1.8\lesssim F/F_c\lesssim 3$), the rate increases like $W_{\rm fOBI}\sim F^{2.66}$ for helium. Also here the growth is steeper than the one found for hydrogen.  

\section{Aspects beyond atomic physics}

\subsection{Application in plasma simulation codes}
Analytical rate formulas for strong-field ionization are useful as
elementary building blocks in numerical codes for simulating of the many-body dynamics 
in laser-generated plasmas. Early implementations in particle-in-cell codes relied on
the ADK formula solely (see, e.g., \cite{Bruhwiler,Nuter,Chen}). In more recent years, 
this approach has been extended in order to properly account for ionization in the OBI 
regime as well. To this end, piecewise combinations of the ADK formula with the fOBI rates 
\eqref{Bauer} and/or \eqref{Kostyukov} have been implemented \cite{Artemenko, Ouatu}.
These rate formulas are 'glued together' at their crossing points to obtain a 
continuous field dependence. Also the nOBI rate \eqref{Tong-Lin} of Tong and Lin 
is applied in particle-in-cell codes \cite{smilei}.

However, the various implementations of analytical field ionization rates presently in use
can lead to different numerical outcomes, as has been analyzed recently \cite{Mironov}.
This happens when the rate formulas are extended beyond their range of applicability. 
For example, the tunneling-like nOBI formula \eqref{Tong-Lin} of Tong and Lin underestimates 
the ionization rate for $F \gtrsim 2.5F_c$, while the linear fOBI formula \eqref{Kostyukov} 
by Kostyukov and Golovanov underestimates the rate for $F > F_a$.

This situation can be improved by the piecewise combined rate formula 
in Eq.~\eqref{full-rate}. It offers the advantage to achieve not only a smooth 
transition from the tunneling regime to the near-OBI domain of Tong and Lin to the 
quadratic far-OBI region of Bauer and Mulser and eventually to the linear far-OBI regime, 
but is additionally in close agreement with fully numerical data. Accordingly, our 
combined rate overcomes the difficulties addressed in \cite{Mironov} and can, thus, be 
useful for application in large-scale plasma simulation codes.

\subsection{Relation to strong-field pair production}

The transition of the ionization rate from the tunneling to the near- and far-OBI regimes has a close analogy in other strong-field phenomena. For example, in the strong-field Breit-Wheeler process, electron-positron pairs are created from vacuum by a high-energy $\gamma$-photon of frequency $\omega'$ colliding with an intense laser field \cite{Ritus}. In the regime of high laser intensities, where the quantum nonlinearity parameter $\chi\sim \omega'F/F_{\rm cr}\ll 1$ is still small, the pair production rate $W_{\rm BW}\sim e^{-8/(3\chi)}$ has a tunneling-like form, very similar to Eq.~\eqref{Landau}. Here $F_{\rm cr}=c_0^3$ denotes the critical field strength of quantum electrodynamics, with the speed of light $c_0 = 137$\,a.u. in vacuum. The tunneling rate formula continues to hold approximately in the intermediate 'near over-barrier' region of $\chi\sim 1$, but overestimates the exact result (see, e.g., Fig.~5 in \cite{Alina}). Eventually, in the 'far over-barrier' regime for $\chi\gg 1$, the field dependence of the rate changes into a fractional power-law of the form $W_{\rm BW}\sim F^{2/3}$. In case of the strong-field Bethe-Heitler process, that is electron-positron pair production in combined Coulomb and laser fields, the exponential tunneling-like rate transitions to an $W_{\rm BH}\sim F\ln(F)$ dependence in the far over-barrier regime \cite{Milstein}. Hence, strong-field ionization and pair production share qualitatively similar, though not identical nonperturbative features.

\section{Conclusion}
Over-barrier ionization of hydrogen and helium atoms has been studied 
on the basis of phenomenological rate formulas. It was shown that the
successful analytical rate expressions by Tong and Lin \cite{TongLin}
and by Zhang et al.~\cite{Zhang} for ionization in a near-OBI regime
can be physically interpreted as arising from modifications of the 
ADK tunneling rate by the Stark effect and a widened range of electron
emission angles. Our corresponding approach provides ionization rates
that are in close agreement with these established empirical formulas and
with fully numerical results for hydrogen and helium. 

We emphasize that our considerations were purely phenomenologically oriented 
and, consequently, do not represent rigorous derivations. They may thus be 
regarded complementary to a recent thorough treatment of (relativistic) 
strong-field ionization and OBI \cite{Klaiber}.

In the far-OBI regime of very large (but still nonrelativistic) field 
strengths we have presented compact four-parameter rate formulas that
agree very well with available numerical data and generalize analytical
expressions obtained previously by Bauer and Mulser \cite{BauerMulser}
and by Kostyukov and Golovanov \cite{Kostyukov}. Our near-OBI and far-OBI 
rate formulas can be joined to cover a very broad range of field strengths. 
The resulting analytical expression can be implemented in large-scale 
laser-plasma codes to describe elementary ionization events.



\begin{thebibliography}{33}
\bibitem{sign} As pointed out in \cite{Kostyukov}, a misprinted 
minus sign needs to be fixed in the exponent of Eq.~(8) in \cite{Zhang}.
\end{thebibliography}
\end{document}